\documentclass[sigconf,nonacm]{acmart}

\AtBeginDocument{}

\usepackage{adjustbox}
\usepackage{hhline}
\usepackage{longtable}
\usepackage{mdframed}
\usepackage{multirow}
\usepackage{pifont}
\usepackage{xspace}
\usepackage{stringstrings}
\usepackage{subcaption}
\usepackage{textcase}
\usepackage{titlecaps}
\usepackage{longtable}

\newcommand{\framework}[0]{SoccerGuard\xspace}
\newcommand{\preprocessingblock}[0]{Preprocessing Block\xspace}
\newcommand{\mlblock}[0]{Automated Machine Learning Block\xspace}
\newcommand{\soccerdashboard}[0]{Soccer Dashboard\xspace}
\newcommand{\soccermon}[0]{SoccerMon\xspace}
\newcommand{\pmsys}[0]{pmSys\xspace}
\newcommand{\gpsdata}[0]{GPS data\xspace}

\newcommand{\matchreports}[0]{WyScout\xspace}

\newcommand{\nmccv}[0]{45\xspace}
\newcommand{\nmodels}[0]{five\xspace}
\newcommand{\nexperiments}[0]{90\xspace}
\newcommand{\nexperimentcomponents}[0]{four\xspace}
\newcommand{\nexperimentsdisplay}[0]{eight\xspace}

\usepackage[acronym]{glossaries}
\newacronym{acwr}{ACWR}{Acute Chronic Work Load}
\newacronym{ai}{AI}{Artificial Intelligence}
\newacronym{ann}{ANN}{Artificial Neural Network}
\newacronym{atl}{ATL}{Acute Training Load}
\newacronym{ctl}{CTL}{Chronic Training Load}
\newacronym{ffrc}{FFRC}{Female Football Research Centre}
\newacronym{uefa}{UEFA}{Union of European Football Associations}
\newacronym{concacaf}{CONCACAF}{Confederation of North, Central America and Caribbean Association Football}
\newacronym{fifa}{FIFA}{F\'ed\'eration Internationale de Football Association}
\newacronym{gps}{GPS}{Global Positioning System}
\newacronym{si}{SI}{Système International}
\newacronym{ioc}{IOC}{International Olympic Committee}
\newacronym{ml}{ML}{Machine Learning}
\newacronym{rpe}{RPE}{Rating of Perceived Exertion}
\newacronym{smdcs}{SMDCS}{Sport Medicine Diagnostic Coding System}
\newacronym{srpe}{sRPE}{Session RPE}
\newacronym{tsc}{TSC}{Time Series Classification}
\newacronym{svc}{SVC}{Support Vector Classifier}
\newacronym{gan}{GAN}{Generative Adversarial Network}
\newacronym{lstm}{LSTM}{Long Short-Term Memory}
\newacronym{auc}{AUC}{Area under the Curve}
\newacronym{roc}{ROC}{Receiver Operating Characteristic}
\newacronym{tpr}{TPR}{True Positive Rate}
\newacronym{tnr}{TNR}{True Negative Rate}
\newacronym{f1}{F1}{F-Score}
\newacronym{host}{HOST}{Department of Holistic Systems}
\newacronym{gui}{GUI}{Graphical User Interface}
\newacronym{mccv}{MCCV}{Monte Carlo cross-validation}

\begin{document}

\title[\framework]{\framework: Investigating Injury Risk Factors for \\
Professional Soccer Players with Machine Learning}

\author{Finn Bartels}
\affiliation{
  \institution{Leipzig University}
  \city{Leipzig}
  \country{Germany}}
\author{Lu Xing}
\affiliation{%
  \institution{University of Oslo}
  \city{Oslo}
  \country{Norway}}
\author{Cise Midoglu}
\affiliation{%
  \institution{Forzasys}
  \city{Oslo}
  \country{Norway}}
\author{Matthias Boeker}
\affiliation{%
  \institution{SimulaMet, OsloMet}
  \city{Oslo}
  \country{Norway}}
\author{Toralf Kirsten}
\affiliation{%
  \institution{University Medical Center Leipzig}
  \city{Leipzig}
  \country{Germany}}
\author{Pål Halvorsen}
\affiliation{%
  \institution{SimulaMet, OsloMet, Forzasys}
  \city{Oslo}
  \country{Norway}}

\renewcommand{\shortauthors}{Bartels et al.}

\begin{abstract}
We present \framework, a novel framework for predicting injuries in women's soccer using \gls{ml}. This framework can ingest data from multiple sources, including subjective wellness and training load reports from players, objective GPS sensor measurements, third-party player statistics, and injury reports verified by medical personnel. We experiment with a number of different settings related to synthetic data generation, input and output window sizes, and \gls{ml} models for prediction. Our results show that, given the right configurations and feature combinations, injury event prediction can be undertaken with considerable accuracy. The optimal results are achieved when input windows are reduced and larger combined output windows are defined, in combination with an ideally balanced data set. The framework also includes a dashboard with a user-friendly \gls{gui} to support interactive analysis and visualization.
\end{abstract}


\begin{teaserfigure}
    \centering
    \includegraphics[width=0.9\columnwidth]{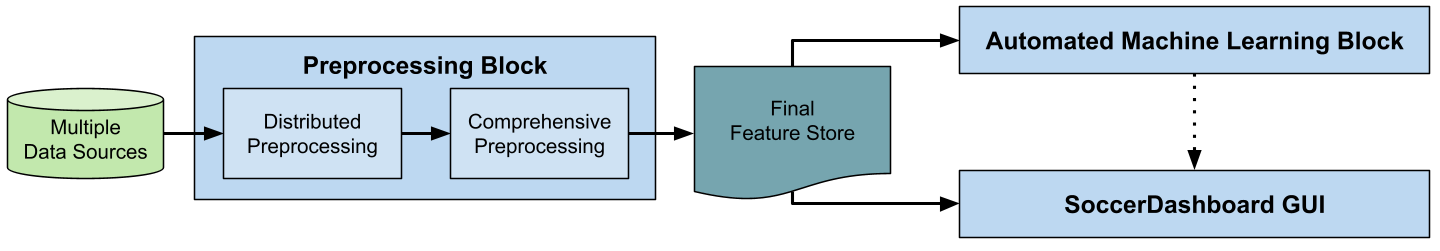}
    \caption{Overview of the \framework framework: Data import from multiple sources, a \MakeLowercase{\preprocessingblock} comprised of two preprocessing blocks, an \MakeLowercase{\mlblock}, and the \soccerdashboard GUI.} 
    \label{fig:framework-overview}
\end{teaserfigure}

\maketitle

\section{Introduction}\label{sec:introduction}

The training intensity of professional soccer players has experienced a tremendous increase in the past decade, in parallel to the growing athleticism in the sport. From the 2006/2007 to the 2012/2013 English Premier League seasons, the mean distances of high-intensity running increased by 28\% with ball possession and by 31\% without ball possession~\cite{barnes2014evolution}. Players performed twice as many sprints during the \gls{uefa} Champions League in the 2018/2019 season compared to prior years~\cite{uefa2018}. With an elevated number of training sessions, matches, and tournaments coupled with a larger number of injuries~\cite{nassis2020increasematches, ekstrand2016hamstring}, the reduction of injury risk factors has become an increasingly important aspect of professional soccer teams~\cite{gabbett2017athlete}. Given the comparatively limited research conducted on female athletes, it is crucial to gain a deeper understanding of the risk factors associated with injuries in female athletes~\cite{bodenner2015}.

In this paper, we make several contributions to the state of the art in injury risk analysis for the domain of women's soccer. The central contribution is the \framework framework consisting of three major pipelines: The \preprocessingblock, the \mlblock, and the \soccerdashboard. The study also includes \nexperiments unique \gls{ml} experiments, each with different settings across \nexperimentcomponents experiment components, focused on predicting an injury event in a \gls{tsc} scenario. Moreover, the \soccerdashboard offers an interactive tool that facilitates a more comprehensive understanding of the characteristics of the athletes' data and potential injury risks. \framework is designed to be flexible in order to facilitate the integration of multiple data sources. The possibilities of monitoring data in soccer clubs and related work, including \gls{ai} in prediction systems in the domain of soccer are demonstrated. With the \gpsdata and \pmsys data included in this study, several studies base their work and results on a similar subset of these instances of data~\cite{van2021machine, rossi2018effective, lovdal2021injury}. It is proposed that the fusion of different data sources can enhance the prediction accuracy of injury detection by~\gls{tsc}.

\section{Background and Related Work}\label{sec:background}

\subsection{Athlete Monitoring}

The monitoring of athletes is a crucial aspect of the decision-making process within sports teams. This concept has applicability in numerous domains within the realm of sports. In the context of soccer, the areas of performance that are typically observed include the measurement of players in matches and competitions, tactics and in-game strategies, the scouting of potentially new players, and negotiations between clubs and between clubs and players. A further significant application is the monitoring of player workload and well-being over the course of a season~\cite{gabbett2017athlete, robertson2017decsupport}.

\textbf{\matchreports:} Specific and very detailed information on teams and players' match performances are provided by services like \matchreports. Such services work with analysts to collect data on a large number of statistical attributes. This includes actions without contact with an opposing player, such as passing or dribbling, and actions with contact with an opposing player, such as fouls committed or suffered by a player, tackles, and several different types of challenges~\cite{wyscout2023datacollection}. \matchreports is a product, provided by Hudl, which is a competitor with one of the highest market shares~\cite{swanson2021hudl}. Besides the analysis of sports data, we can see approaches to increase the value of the data that is available for teams. Since the extensive use of \gls{ml} in various fields, this technology is used in the research of sports and in the environment of sports clubs~\cite{blom2019using}.

\textbf{\pmsys:} The \pmsys athlete monitoring system is a discontinued performance monitoring system that enabled athletes to subjectively evaluate their individual training load and wellness behavior on a regular basis~\cite{pmsys_web}. \pmsys was provided by the \gls{ffrc}, a collaboration of the UiT - The Arctic University of Norway (Universiteit van Troms{\o}) and the Department of Holistic Systems of Simula. The athletes answered daily questions about their training load, wellness, and injuries. This led to an overall view of the well-being of each player. The coaching staff can monitor the results of their athletes' assessments in a web application, which includes plots of the team or individual athletes~\cite{hoang2015pmsys}.

\subsection{Injury Analysis, Prediction, and Prevention}

An already efficient and established approach to reduce injuries in the domain of soccer is the \gls{fifa} 11+ program~\cite{sadigursky2017fifa}. This program is a written manual complemented with images of exercises. It was designed to include a number of warm-up exercises to target the prevention of injuries in training sessions and matches. In a study with 6,344 players, the \gls{fifa} 11+ program reduced the number of injuries by 30\% in the intervention group compared to the control group. Programs like \gls{fifa} 11+ are not designed to manage the load of players but provide teams with standardized exercises to reduce the risk of injuries independent of the loads of individual players.

\textbf{Use of AI/ML:} Applying \gls{ml} to predict the injury risk is geared towards the situation of each individual athlete, backed by their data. The last years have seen increasing use of \gls{ai} in predicting parameters in various sports~\cite{mccullagh2013anninj}. Among these are the performance prediction, match result prediction, and injury risk prediction in different disciplines of sports like soccer, American football, Australian football, basketball, and baseball~\cite{carey2018predictive, bunker2022application, igiri2014improved, claudino2019current, van2021machine}. Particularly since the 2010s, the majority of studies in a systematic review related to injury risk prediction in team sports referred to the use of \gls{ai} techniques as an option to provide better results than traditional statistical methods~\cite{claudino2019current}. In particular, these studies are mainly based on soccer data. The techniques used on the soccer data were mainly \gls{ann} and Decision Tree Classifier. But also, the selected studies of other sports focused mainly on these methods besides \gls{svc}~\cite{claudino2019current}.

\textbf{Injuries in women's soccer:} Given that female athletes are more likely to suffer from different injuries than male athletes, there is a clear need for an open-source injury data set and studies on this~\cite{giugliano2007acl}.
Open resources can contribute to a better understanding of injury risks in female athletes~\cite{bodenner2015}.
Therefore, in the following, we rely on \soccermon as a comprehensive data set of multiple types of data~\cite{midoglu2024soccermon}. It is the main data source that was used to build the \framework.

\textbf{Visualization of injuries:} Authors in~\cite{boeker2023soccer} present a Streamlit for the analysis of parts of the subjective data from the \soccermon dataset (using wellness and training load reports), but do not consider injuries. The dashboard also does not use the objective GPS data from the \soccermon dataset, or any external data sources, such as player statistics or match information.

\section{Datasets}\label{section:datasets}

In this study, we use a comprehensive dataset comprising a total of 114 attributes and 4448 data points. The data points were collected from the activities of 37 players between February 2020 and November 2021. This period encompasses two seasons and the corresponding preparation phase. The data set comprises 322 days of training or competitive matches.

\subsection{\soccermon}

\soccermon is the largest extant open-source soccer athlete data set, comprising subjective as well as objective data, or more precisely, GPS reports, from a two-year period. The data set comprises exclusively data of female athletes. Subjective data comes from \pmsys. Teams representing Toppserien already started utilizing the PmSys system in the season of 2020. This allows for the examination of data over a comprehensive time period of one year, encompassing the years 2020 and 2021.  The \gpsdata is derived from the tracking of athletes during training sessions and competitive matches. The \soccermon data set is open access and available on Zenodo. It includes 33,849 subjective reported parameters and 10,075 objective parameters~\cite{midoglu2024soccermon}.

\textbf{Subjective reports:} The subjective data on which the players are assessed is divided into five groups of metrics. The metric groups considered in this work are training load assessments and wellness assessments. Further assessments include injury, illness, and subjective game performance reports. The data contained in the game performance reports was deemed irrelevant for the purposes of the analysis of injuries, and the manual injury and illness reports by athletes lacked missing data. Consequently, those categories were not considered in the \framework. The training load is quantified using 11 distinct features. In addition to the specified duration of subjective training, the players are also asked to provide their \gls{rpe} on a scale from zero to ten. The \gls{rpe} describes the perceived exhaustion after a training session or match. Subsequently, nine additional attributes are derived from the duration and \gls{rpe}, which serve to indicate the training load. Those attributes include \gls{srpe}, \gls{ctl} of 28 days and 42 days, daily load, weekly load, \gls{atl}, monotony, strain and \gls{acwr}. The efficacy of \gls{srpe} has been substantiated by a multitude of studies~\cite{haddad2017srpe}. This constitutes the utilization of this attribute in conjunction with the additional derived attributes. The wellness assessment consists of seven attributes. This constitutes the utilization of this attribute in conjunction with the additional derived attributes. The attributes for assessing the players' state of wellness are rated with the different attributes of fatigue, mood, readiness, soreness, stress, duration of sleep, and quality of sleep. These factors indicate the perceived degree of physical and mental exhaustion of the players, which also includes the state of a player's wellness outside of training sessions. The importance of sleep can be recognized, among other things, by the fact that the \gls{ioc} named sleep as an influencing factor in the risk of injury in their consecutive paper about developing healthy youth athletes in 2015~\cite{walker2017we, bergeron2015international}. Sleep quality is characterized by the frequency of sleep phases and the duration of the individual sleep phases, which may affect indicators like feeling well-rested after sleep, the ease of waking up, and having headache or other symptoms~\cite{walker2017we, krystal2008sleepq, yi2006sleepq}.

\textbf{\gpsdata:} The objective part of the \soccermon dataset comes from GPS sensor measurements that were collected with a frequency between 10Hz-100Hz during entire sessions. It was collected using the StatSports APEX Athlete Series System~\cite{statsports_web}. The system comprises a vest worn by the athletes, which is equipped with a GPS performance tracker. The aggregation of data results in the description of a training session for a player, which is represented by a number of attributes. 16 key metrics are evaluated during the performance of a player. Some of the attributes are intended for the evaluation of measurement inaccuracies, such as the number of satellites available for the measurement.

\subsection{\matchreports Reports}

\matchreports is a product developed by Hudl that provides professional soccer teams and participants with a comprehensive library of soccer statistics. The provision of match statistics reports has enabled the further enrichment of the data extracted from \soccermon. It is important to note that the available data at Simula \gls{host} is only available for matches in the 2021 season. However, \soccermon covers training sessions from the seasons of 2020 and 2021. Taking this into account, it corresponds to a total of 18 games in the entire 2021 season. The reports consist of a large number of attributes. The comprehensive data set encompasses 38 attributes from \matchreports, collectively providing insight into the performance and play style of each individual player in a given match. A total of 235 match entries from 19 different players are available for the 2021 season. The data set includes a number of attributes that represent offensive actions (e.g., goals, expected goals, and assists) and defensive actions, as well as challenges that indicate physical contests (e.g., tackles, fouls, and challenges).

\subsection{Injury Reports}

The players were able to report injuries and illnesses in the daily \pmsys assessment. The number of injuries players suffered suggests that only a small number of the injuries were reported in the assessments through the app. This has been confirmed by the team staff. The medical therapists of the competing teams provided reports on the injuries sustained by their players. The report comprises a series of attributes that provide a detailed account of the incidents in question. A total of 43 injuries have been recorded in 18 players. The report made a distinction in the cause, activity, and area of the injuries. These characteristics, as well as the target variable of an injury event, were incorporated into the comprehensive data set. The findings of the area attribute analysis were used to derive and add information on the body region as an attribute~\cite{page2012bodyregions}.

\subsection{Synthetic Data}

Given the relatively small number of 4448 general data points in the data set (less than 1\% of which include an injury, with 43 data points in total), it was deemed appropriate to utilize synthetic data in order to assess the impact on performance. Prior to the training process, the data sets are segregated into two distinct categories: Training data and test data. The generation of synthetic data is then conducted on the basis of the training data~\cite{patki2016sdv}. The detailed procedure is described in section \ref{sec:framework}. 

\section{Proposed Framework}\label{sec:framework}

We propose the \framework framework, a comprehensive framework which aims to address the research objective introduced in section~\ref{sec:introduction}. \framework consists of the \preprocessingblock, the \mlblock, and the \soccerdashboard. The preprocessing and \gls{ml} pipelines are designed with a structure oriented towards MLOps and CRISP-DM~\cite{google_mlops}. SRP-CRISP-DM is an adapted framework in the field of sports prediction, and more specifically in the field of sports result prediction~\cite{bunker2019srpcrisp}. Figure~\ref{fig:framework-overview} presents an overview of the \framework framework, depicting the \preprocessingblock, the \mlblock and the \soccerdashboard. The third part of the framework is a \gls{gui} built as a web application which visualizes various aspects of the injury analysis and prediction which are part of the \mlblock. The web application, called \soccerdashboard is implemented with Streamlit.

\begin{figure}
    \centering
    \includegraphics[width=0.9\columnwidth]{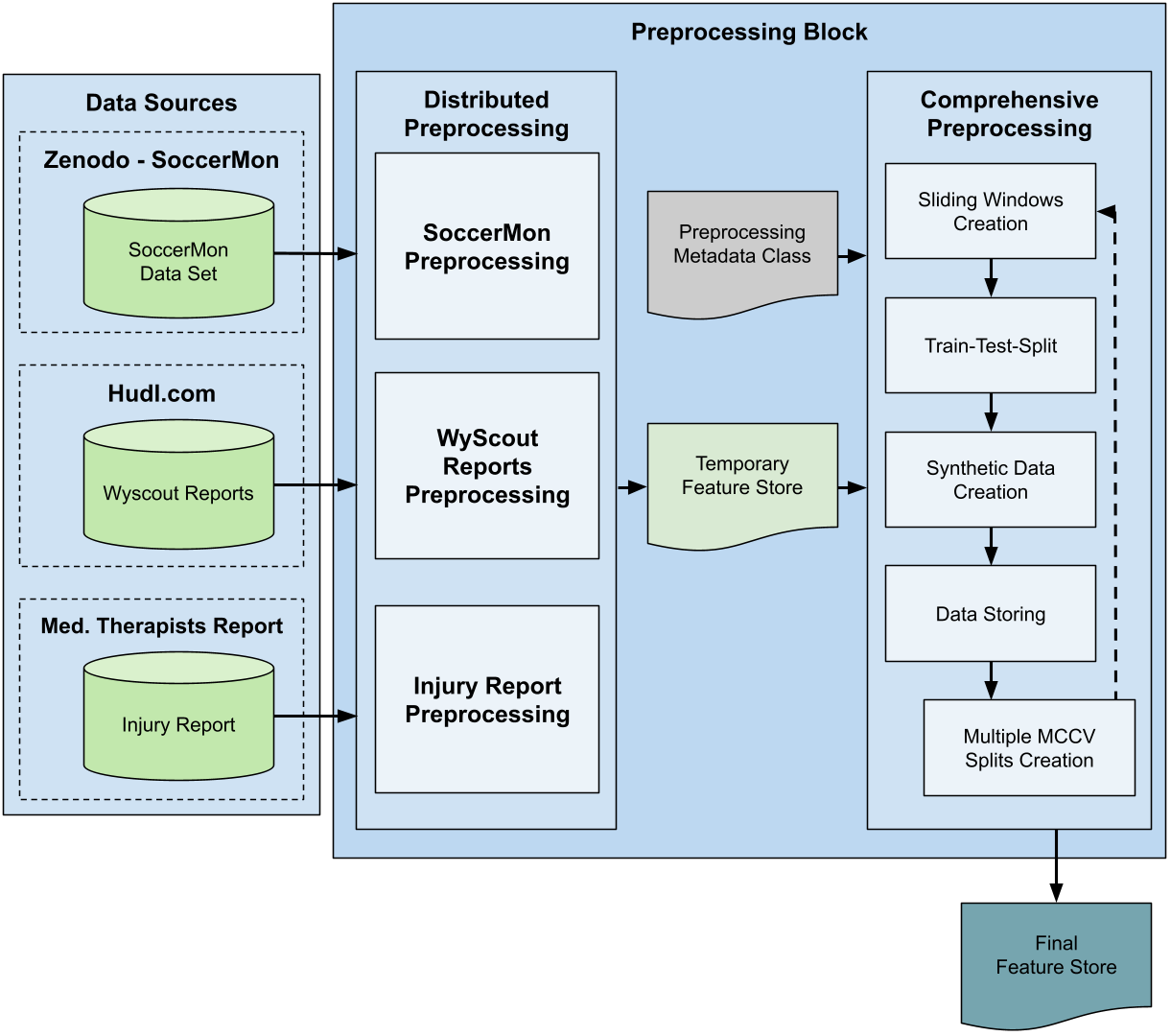}
    \caption{The \MakeLowercase{\preprocessingblock} is comprised of two distinct preprocessing blocks. The first is the distributed preprocessing of the raw data, and the second is a comprehensive preprocessing of the temporary feature store.}
    \label{fig:preprocessing}
\end{figure}

\subsection{\preprocessingblock}

\subsubsection{Data Source and Feature Selection}
The data source and feature selection is the initial step of the \framework framework process. In general, there are three data sources considered in this study, namely the \soccermon data set, the \matchreports reports, and the injury reports. When selecting features from the \soccermon dataset, we distinguish between two sources, subjective data and objective data (i.e., \gls{gps} sensor data). Starting with the subjective data, this part of \soccermon was stored in an approach of .csv files with each of the subjective features separately, including player IDs and dates. In each of the files, a column represents a unique player ID. The \gpsdata is provided in separate daily .zip files. Each one contains a parquet file of a single player within a given session. These files are stored in monthly and yearly folders. An initial loading script was implemented to concatenate the session files of all players into a monthly structure of parquet files. This resulted in a larger single parquet file for each month between June 2020 and November 2021. The \pmsys enables players to put in information regarding their injuries into their daily reports. However, upon observation of the analyzed team's reports, it became evident that this process was not consistently followed. This means, only a small number of injuries was considered in the daily reports. To address this issue, manual reports from the team's medical therapists were obtained afterward to fill the gap of missing injury records.

\subsubsection{Distributed Preprocessing}

The distributed preprocessing of \soccermon data is comprised of two distinct pipeline components. Each for a separate stream of data: Subjective \soccermon data and \gpsdata are handled separately. In both cases, a functionality to process raw data is provided, which is separated in teams, in teams and year, or in teams and month. For the processing of \gpsdata, several functions were applied to determine if there are implausible values in the \gpsdata that originate from signal quality indicators and geographical data variables. Thresholds were defined in consultation with StatSports and according to criteria defined by the Simula Institute. Around 92.91\% of the data points were retained, which is a total of 2.59 billion data points over 18 months. In this first step, \gpsdata sets were saved monthly. The subjective data did not require significant processing in the initial stage. The medical therapist's report, which contains the target variable, provides various information about the injuries. In addition to the target variable, various features pertaining to the injuries are provided, which are specific to each team. The existing infrastructure for mapping injury reports to the main data sources enables the analysis and modeling to be extended to more specific types of injury in further trials. This process entails overcoming the anonymity of the participants and matching their names according to the Levenshtein distance~\cite{berger2020levenshtein}.
    
In the subsequent phase, the attributes in the \gpsdata set were augmented by the derivation of additional attributes and the aggregation of attributes on a daily basis. From a micro perspective, the aggregation of data points that are within a second is included. The derivation of new features involved multiple significant steps. The first step was the application of the Haversine formula to measure the distance between data points geographically. Furthermore, the speed was converted to \gls{si} units ($\frac{m}{s}$). The derivation step also encompasses the construction of time in seconds and the delineation of players' positions within heart rate and speed zones. Any derived attributes were already mentioned in section~\ref{section:datasets}. Finally, the integration of \gpsdata, subjective data, and injury data was conducted using the attributes \textit{date} and \textit{player ID}. The format of the resulting data set is a comprehensive data set of data points representing a tracked training session or a competitive match. Each data point is derived from \gpsdata, including optional subjective assessment data and match statistics from \matchreports.

\subsubsection{Comprehensive Preprocessing}

The objective of the comprehensive preprocessing block is to prepare a data set for use in a specific injury prediction context. Missing values are substituted within one of a number of univariate or multivariate techniques. Numeric values can be substituted with the median, a linear interpolation or an iterative imputation for a player. Categorical values are substituted with a label of 'unknown' value. In this block, the specific features to be utilized in the construction of the \gls{ml} model are selected. Users feed the script to run this block with information on: Numeric imputation technique, usage and number of synthetic data points, window size and proportion of the test data and a \gls{mccv} instance. 

In the main part of the block, a data set is generated which was built by running a sliding window on the original data. The size and limitations of a sliding window were predefined. A sliding window represents a number of sessions, training as well as matches, in a certain time for a player. This data set is already split into training and testing. Optionally a user can increase the number of training data with generated synthetic data. This is of both types, injury and non-injury events. The Synthetic Data Vault library was employed for the optional generation of synthetic data~\cite{patki2016sdv}. The synthesizer generates synthetic data using a trained \gls{gan}. The real data of training and testing, as well as the optional synthetic data for training, are organized in a folder structured form for further pipeline blocks.

A further optional function determines the number of random cross-validation folds to be created. This is done using the \gls{mccv} method~\cite{shao1993mccv}. It is a potential solution that may be beneficial in cases where the overall data size is relatively small.

\begin{figure}
    \centering
    \includegraphics[width=0.9\columnwidth]{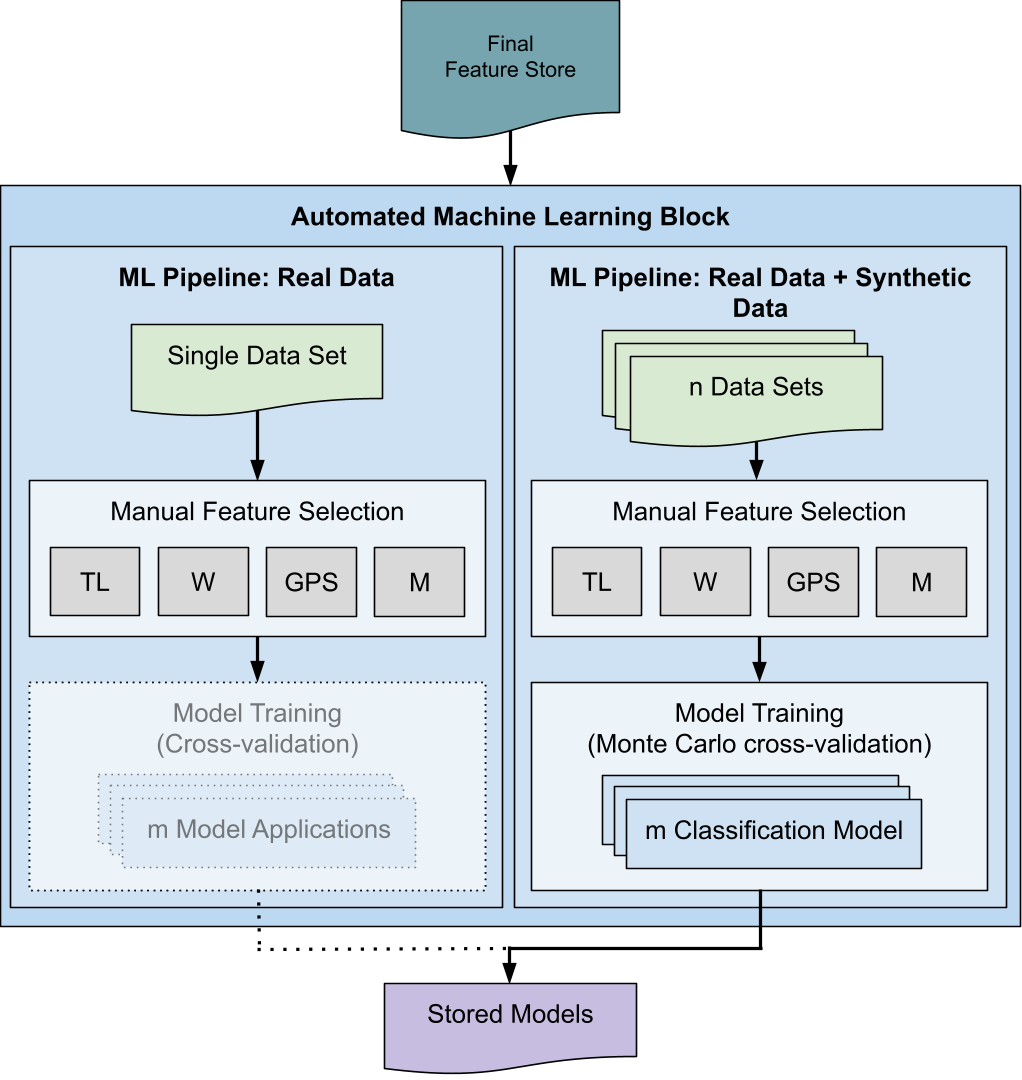}
    \caption{The \MakeLowercase{\mlblock} consisting of two pipelines for training individual models on real data and training on both real and synthetic data.}
  \label{fig:ml}
\end{figure}

\subsection{\mlblock}\label{sec:framework:sub:ml}

The automated machine learning block enables the differentiation between the modeling of real data exclusively and that of real data in conjunction with synthetic data. The block has been constructed with the intention of enabling the training of a number of different \gls{ml} approaches on a similar infrastructure. The training data is divided according to the model being trained, resulting in a training dataset and an optional validation dataset. Normalization of each data set is done separately using a standard scaler. The actual framework provides five different \gls{ml} approaches to be trained on. Each model is applied on the predefined data points of window size. The aforementioned structure combines each feature with the number of sessions considered and the target variable (e.g. \textit{speed\_km\_h\_max\_1, speed\_km\_h\_max\_2, ..., speed\_km\_h\_max\_n, ..., injury}). A number of basic approaches like Logistic Regression, Random Forest, and \gls{svc} are supplemented by further applications~\cite{carey2018predictive}. More sophisticated models in this block are XGBoost and a \gls{lstm}. The application of XGBoost is a common approach in the field of event prediction, including in the context of \gls{tsc}~\cite{lovdal2021injury, bunker2022application}. Also \gls{lstm} neural networks are a well-established and frequently employed approach~\cite{wiik2019pred_lstm}. 

The overall performance of an approach was validated using several accuracy metrics of an attempt. For the approaches including synthetic data, each model was trained \nmccv times as well as tested \nmccv times. This approach was chosen because it was not possible to perform cross-validation with a small number of sliding window data points. The final result was drawn as the mean of the \nmccv resulting values. \gls{mccv} was employed to enhance the reliability of the results, given the limited number of data points in the real data set. A set of \nmccv CV rounds was selected~\cite{shan2022mccv}. This is equivalent to a 10-fold cross-validation, with 8 folds in training and 2 folds in testing. Such a high number of rounds was chosen to obtain the most meaningful and valid result possible, in accordance with the feasibility and cost of the training.

\begin{figure}
    \centering
    \includegraphics[width=0.9\columnwidth]{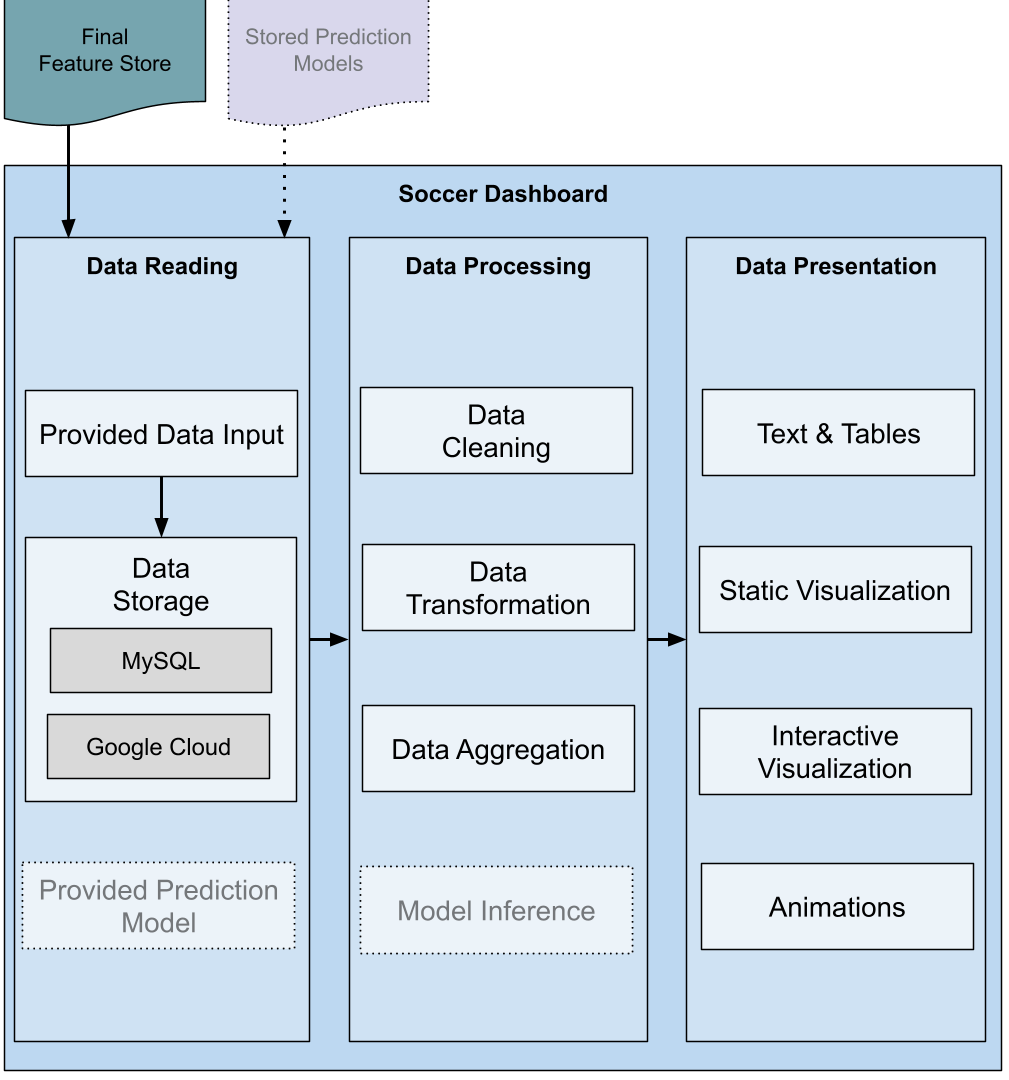}
    \caption{Underlying pipeline of the \gls{gui}/web application: \soccerdashboard.}
  \label{fig:gui-pipeline}
\end{figure}

\begin{figure*}
    \centering
    \frame{\includegraphics[width=\textwidth]{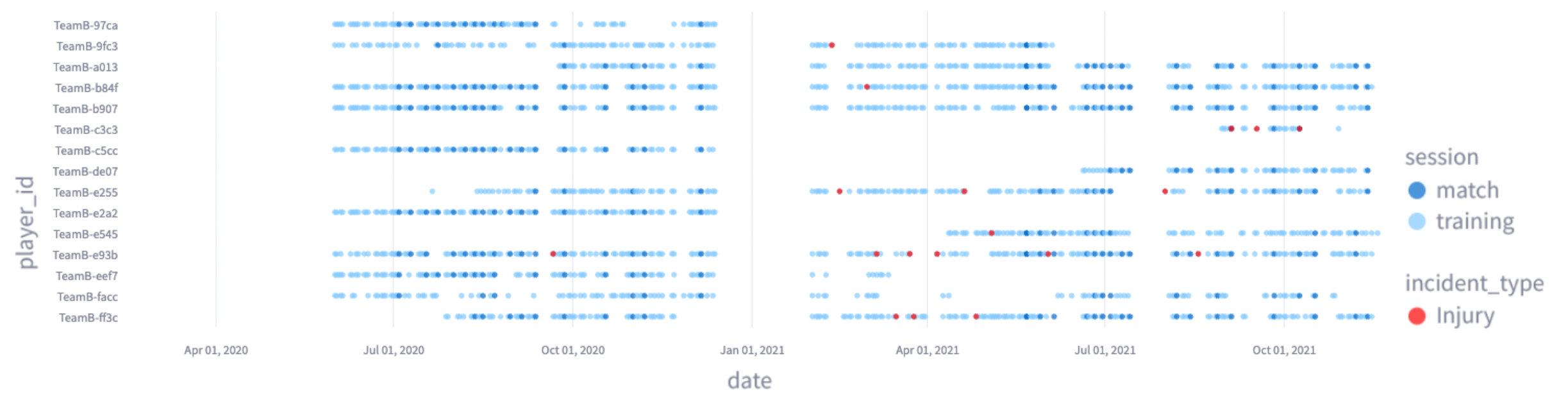}}
    \caption{\soccerdashboard: An overview of the matches, training sessions and injuries by date and player for the entire team.}
  \label{fig:gui-sample1}
\end{figure*}

\begin{figure}
    \centering
    \frame{\includegraphics[width=\columnwidth]{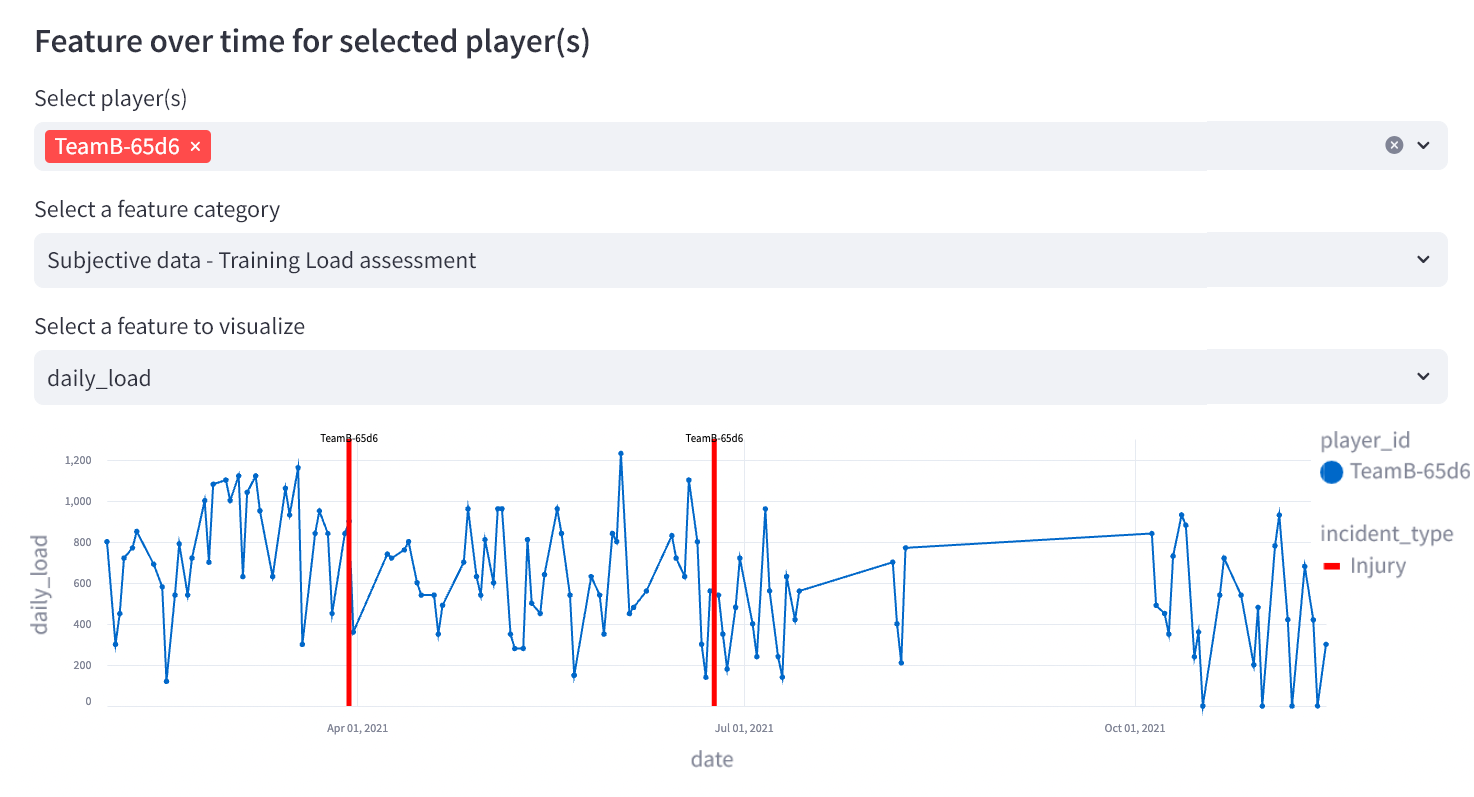}}
    \caption{\soccerdashboard: A detailed view on the trend of a selected feature for a selected player including highlighted injuries by date.}
  \label{fig:gui-sample2}
\end{figure}

\subsection{\acrfull{gui}}

The third component of the \framework framework is a \gls{gui} built as a web application using Streamlit, called \soccerdashboard. \soccerdashboard is designed to visualize the data, the injury analysis, and the prediction results.
The architecture of the \soccerdashboard is shown in Figure~\ref{fig:gui-pipeline}. \soccerdashboard consists of three main parts: Data reading, data processing, and data presentation. The data reading part is responsible for reading the data from the data source. This part is able to read the data from a wide range of data sources, such as MySQL database, Google Cloud Storage, and local files. It also supports reading data from different file formats, such as CSV, JSON, and Parquet. After being read, the data is processed in the data processing part to make it ready for visualization. The data processing part includes data cleaning, data transformation, and data aggregation. The data cleaning part is responsible for cleaning the data by filtering out missing values, duplicates, and other data that is not needed or not suitable for visualization. The data transformation part is responsible for transforming the data into a format that is suitable for visualization, such as restructuring the data, joining multiple data sources, and converting data types. The data aggregation part is responsible for aggregating the data into a format that is suitable for visualization, such as grouping the data by certain attributes, summarizing the data, and calculating statistics. Finally, the data is presented in the data presentation part. The data presentation part includes various types of visualizations, such as tables, charts, and animations. The data presentation part is designed to be interactive, allowing users to explore the data in different ways, such as filtering the data, sorting the data, and drilling down into the data.

\soccerdashboard is designed to be user-friendly, intuitive, and easy to use, allowing users to quickly and easily explore the data, analyze the data, and make informed decisions based on the data. The \soccerdashboard is also designed to be flexible and extensible, supporting a wide range of data sources, data formats, and visualization types. Here, the \soccerdashboard is used to read the data from the final comprehensive feature store, which is the output of the \preprocessingblock, and present the data in an interactive and user-friendly way.

Figure~\ref{fig:gui-sample1} shows a sample screenshot of the \soccerdashboard, which is an overview of all the sessions in the final feature store along with the injury events. Figure~\ref{fig:gui-sample2} shows another sample screenshot of the \soccerdashboard, which demonstrates the ability to explore any feature from the final feature store from one or more players against the target variable, which is the injury event. The \soccerdashboard can automatically detect the data type of each feature and present the data in an appropriate way, such as a line chart for numerical features and a dot plot for categorical features.

We plan to integrate the \soccerdashboard with the \mlblock to visualize the prediction results and compare the performance of different machine learning models. The \soccerdashboard will be able to load the models trained in the \mlblock and make predictions on new data, allowing users to play with the models in an interactive way. It will be possible to make predictions on new data, explore the prediction results, and compare the performance of different models.

\section{Experiments and Results}\label{sec:evaluation}

This study includes a number of experiments to be performed, including the validation of different \gls{ml} models, sliding window sizes and the composition of the original data set. As mentioned in section~\ref{sec:framework:sub:ml}, for each individual experiment run, the average result of each metric was taken from a number of \nmccv runs. Including a set of different values as input size as well as output size, values of event proportion and \gls{ml} models, the number of single experiments is at \nexperiments. Multiplying the number of \nmccv \gls{mccv} rounds to this, a total of 4,050 model instances were trained.

\begin{figure}
  
    \begin{subfigure}[t]{0.9\columnwidth}
        \centering
        \includegraphics[width=\textwidth]{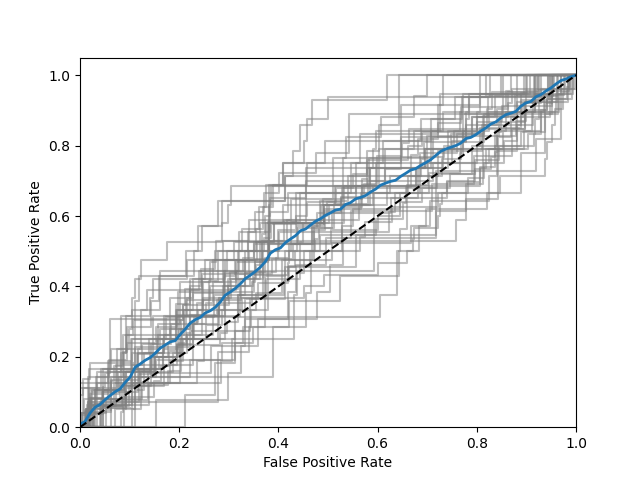}
        \caption{\gls{roc} curve of the \gls{ml} trainings of an \gls{lstm}. \gls{lstm} with an input window of 5 and an output window of 1 and 10\% injury event proportion.}
    \end{subfigure}%
    \vfill
    \begin{subfigure}[t]{0.9\columnwidth}
        \centering
        \includegraphics[width=\textwidth]{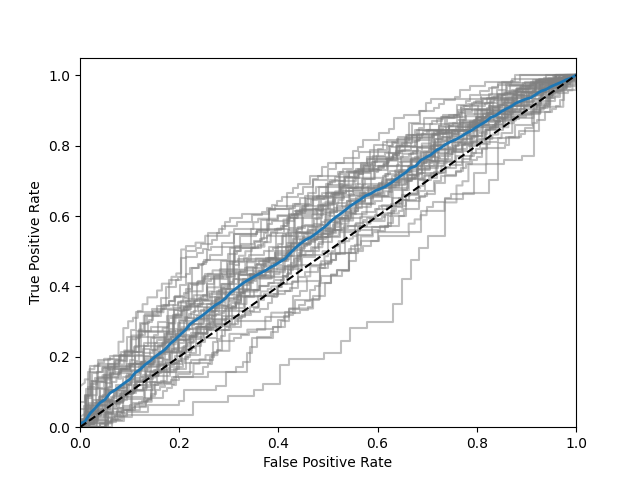}
        \caption{\gls{roc} curve of the \gls{ml} trainings of an \gls{lstm}. \gls{lstm} with input window of 7, an output window of 7 and 50\% injury event proportion.}
    \end{subfigure}
    
    \caption{\gls{roc} curves.}
    \label{fig:ROC_combined}
  
\end{figure}

\begin{table*}

    \small
    
    \begin{tabular}{|c|c|c|c|c|c|c|c|c|c|c|}
    \hline
    
    \multirow{2}{*}{\textbf{ID}}
    & \multicolumn{5}{c|}{\textbf{Combinations}}
    & \multicolumn{5}{c|}{\textbf{Metrics}}
    \\ \hhline{~----------}
    
    & \textbf{Data}
    & \textbf{Event}
    & \textbf{Input}
    & \textbf{Output}
    & \textbf{Features}
    & \textbf{Model}
    & \textbf{Prec}
    & \textbf{TPR}
    & \textbf{F1}
    & \textbf{AUC} 
    \\ \hline \hline 

    I-26 & R+S & 0.25 & 5.0 & 7.0 & TL,W,GPS & logit & 0.341 & 0.042 & 0.07 & 0.559 \\ \hline
    I-56 & R+S & 0.5 & 3.0 & 7.0 & TL,W,GPS & logit & 0.529 & 0.037 & 0.066 & 0.533 \\ \hline
    I-57 & R+S & 0.5 & 3.0 & 7.0 & TL,W,GPS & lstm & 0.532 & 0.05 & 0.089 & 0.497 \\ \hline
    I-58 & R+S & 0.5 & 3.0 & 7.0 & TL,W,GPS & randomforest & 0.623 & 0.014 & 0.027 & 0.618 \\ \hline
    I-70 & R+S & 0.5 & 5.0 & 3.0 & TL,W,GPS & xgboost & 0.821 & 0.006 & 0.012 & 0.595 \\ \hline
    I-71 & R+S & 0.5 & 5.0 & 7.0 & TL,W,GPS & logit & 0.637 & 0.062 & 0.107 & 0.548 \\ \hline
    I-72 & R+S & 0.5 & 5.0 & 7.0 & TL,W,GPS & lstm & 0.561 & 0.066 & 0.113 & 0.533 \\ \hline
    I-75 & R+S & 0.5 & 5.0 & 7.0 & TL,W,GPS & xgboost & 0.706 & 0.022 & 0.04 & 0.578 \\
        
        \hline
    
    \end{tabular}
    
    \caption{Experiment scenarios with respect to \textit{Combinations} of \textit{Data} (R: Real, S: Synthetic), \textit{Event} (proportion of event data), \textit{Input} (Window), \textit{Output} (Window), \textit{Features} (TL: Training load, W: Wellness, GPS: GPS derived data), \textit{Model} (logit: Logistic regression, random forest, svc: Support vector classifier, xgboost, lstm: long short term memory), and resulting \textit{Metrics} (Precision, TPR: True positive rate, F1: F-Score, AUC: Area under the ROC curve).}
    \label{tables:experiment-scenarios}

\end{table*}

\subsection{Evaluation Metrics}

As evaluation metrics, we consider the \gls{auc} of \gls{roc} curve, which is a well-established metric~\cite{ayala2019hamstrinj, lovdal2021injury, wiens2012eval}. In the context of injury prediction in sports, \gls{auc} is a higly represented metric~\cite{van2021machine}. In addition, a number of metrics derived from the confusion matrix were used. This includes the precision, the recall or \gls{tpr}, the \gls{tnr} and the \gls{f1} of a model~\cite{lovdal2021injury, wiens2012eval, ruiz2021eval}. In any considered metric derived from the confusion matrix, the threshold for the classification of a data point is a critical component. As the study did not include an additional measurement of the optimal threshold, a threshold value of 0.5 was considered, with values greater than 0.5 as a predicted injury events.

Another critical component is the proportion of classes. Since the number of injury events is highly underrepresented in the set of real data, oversampling with synthetic data points was a method to compensate this issue. The study includes an experiment designed exclusively for investigating problems in the question of class imbalance. Underrepresentation of an event class can increase the risk to misclassify that event class~\cite{santoso2017synthetic}.

\subsection{Experimental Setup}

The aforementioned classes of experiments are employed in order to test three different window sizes for input and for the output, three different sizes of event proportion and \nmodels models. \gls{tsc} is a method that employs a series of previous events to predict an impending occurrence. The quantity of information utilized in this process can be a determining factor in the information passed to the model and in the accuracy of the outcome. The input window size was set to 3, 5 and 7 sessions respectively. Furthermore, the maximum allowable interval between the initial and final sessions within an input window is 14 days. Input windows of a larger time span were not considered. The difference of days and sessions is that session did not occur on every day as players had resting days in between.

A similar approach was used for the span of output windows. A distinction is made between 1, 3 and 7 sessions for the definition of the output size. In each scenario, the output to be predicted is a single value. In the scenario that the output size is 1, the injury must occur in the subsequent session in order for it to be classified as such. In the event that an output window exceeds a size of 1, an injury event is identified if it occurs within a the time frame of the output window. The maximum allowable interval between the last data point of the input window and the final session of the output window is 14 days, exactly as in the previous procedure.

An additional distinction was made in the event proportion. In the context of the event proportion a distinction between a balanced proportion of data points with a target of injury events and non-injury events and two approaches with unbalanced proportions. As the proportion of injuries in the original data set was less than 1\%, an oversampling was initiated to avoid the wrong classification of the injury events caused by the disproportion. The approach of unbalanced proportion includes 25\% and 50\% of injury events.

In the approaches of models, \nmodels classifiers were used. Those are namely Logistic Regression, Random Forest, XGBoost, \gls{svc} and \gls{lstm}. The objective of this study is not to engage in hyperparameter tuning but the aforementioned options in the experiments.

Table~\ref{tables:experiment-scenarios} presents the list of all \nexperimentsdisplay scenarios that are mentioned in the experiments of sections \ref{ExA}-\ref{ExD}. 
The \gls{roc} curves in figure~\ref{fig:ROC_combined} visualize the differences between the parameters of training that were provided in the scenarios. The gray \gls{roc} curves describe an instance of a single round of \gls{mccv} each, while the blue curve describes the average \gls{roc} curves over \nmccv \gls{mccv} rounds. A comparison of the two figures reveals that the training with an output window of 7 produces smoother curves than the figure with an output window of 1. This is due to a larger number of injury events in the test set. Furthermore, the variance appears to be less extensive in the training with an output window of 7 in comparison to the training with an output window of 1.

\subsection{Experiment A: Injury Event Proportion}\label{ExA}

An appropriate event proportion is necessary in this classification task. We noticed a poor accuracy when applying an event proportion of 0.1 in preliminary tests. The differentiation of event proportions shows that the average \gls{auc} exhibits minimal variation overall. However, a pronounced impact of event proportion is evident in the \gls{f1}, precision, and \gls{tpr}. While precision and \gls{tpr} reach a peak at an event ratio of 0.5. This effect is observable in the comparisons of scenarios I-26 and I-71. 

\textbf{Takeaway:} Upsampling to a larger event proportion is required to achieve an appropriate accuracy. We therefore worked with an event ratio of 0.25 or 0.5.

\subsection{Experiment B: Input Window Size}\label{ExB}

Modifications to the input window result in clear alterations to the \gls{f1}, the \gls{tpr}, the precision, and the \gls{auc} of average results over all \nexperiments scenarios. The percentage change in \gls{f1} between an input window of 3 sessions and an input window of 5 sessions is 40.91\%, while the percentage change between an input window of 3 sessions and an input window of 7 sessions is 27.27\%. An example of such behavior is observable in scenarios I-57 and I-72. 

\textbf{Takeaway:} The findings indicate that utilizing an input window of 5 may yield enhanced outcomes in the prediction of injury events. A larger input window of 7 tends to be better than a small input window of 3.

\subsection{Experiment C: Output Window Size}\label{ExC}

Modifications to the input window yield results that are supplementary to those observed in \ref{ExB}. The percentage change in \gls{f1} between an output window of 1 session and an output window of 3 sessions is 90.91\%, while the percentage change between an output window of 1 session and an output window of 7 sessions is 336.36\%. One illustrative example is the comparison of I-70 and I-75. 

\textbf{Takeaway:} It is recommended that the size of the output window be increased from predicting an exact session event to predicting an event occurring in a range of 5 to 7 sessions.

\subsection{Experiment D: Model Performance}\label{ExD}

Although the \gls{auc} is highest in average results of Random Forest (I-58), the \gls{f1} is significantly higher in Logistic Regression (I-56) and in \gls{lstm} (I-57). A detailed examination reveals that the number of true positive classifications in the Random Forest models is relatively low, whereas the Logistic Regression models demonstrate the highest number of true positive classified events. XGBoost on average represent a mediocre precision, \gls{tpr} and \gls{f1}. 

\textbf{Takeaway:} In the absence of hyperparameter tuning, the methodology of Logistic Regression followed by \gls{lstm} is recommended for the classification of injury risk in a given time series.

\section{Discussion}\label{sec:discussion}

\subsection{Use Cases and Applications}

As the various components of the framework were outlined, it became evident that the use cases and user groups in the domain of soccer and athlete monitoring are quite diverse. Players are able to work with the athlete monitoring to feed the data set of the \framework framework with valuable data. As a result, they are able to gain insights into their own training behavior and load from an objective perspective. It would be advantageous for medical therapists to have a comprehensive view of the predicted injury risks in order to adopt a more sensitive approach when working with athletes. The capability of the \framework framework to facilitate analysis and predictions on injury risks represents an effective method of providing assistance and fostering a symbiosis with the specialized knowledge of the medical therapists. The trainer team of a club will be provided with a stack of tools to perform load management based on the results of the \mlblock and monitoring trends in injury risk factors by using the \soccerdashboard. The \framework allows us to fall back on a number of actions to individualize training sessions for different groups of players. This also includes rotations within the team in cases where factors of the players' load increase the risk of long-lasting injuries.

\subsection{Limitations and Future Work}

The objective was to develop a framework to facilitate the creation of predictive models for injury risk assessment in athletes. The number of usable data points was limited, as was the quite smaller number of injury events. The issue of the limited number of injury events prompted to consider the use of synthetic data generation. Providing a setting to be able to compare different \gls{ml} models on the composed data did not consider hyperparameter tuning. This will be addressed in future work. Furthermore, future work will include the addition of experiments. This will allow for comparing the results generated from the real data set with those generated from the same data set combined with synthetic data. The impact of diverse data sources will be incorporated into the future implementation. The integration of the \mlblock into the \soccerdashboard represents a further stage of development and will facilitate the stakeholders of \framework with valuable predictions.

\section{Conclusion}\label{sec:conclusion}

The \framework introduced a comprehensive and powerful tool for accelerating insights into injury risk factors and predicting injury events. Following the analysis of the experiments and the comparison of the scenarios, a number of best practices were defined in terms of the underlying task and data set. It was determined that achieving an appropriate balance between injury events and non-injury events is essential for the predictive performance of the models. Additionally, it was found that smaller input windows and larger output windows result in better results.

\balance
\bibliographystyle{ACM-Reference-Format}
\bibliography{references}

\end{document}